\documentstyle[12pt,equations,epsf]{article}
\newcommand{\nc}{\newcommand}
\nc{\fdiag}{0}
\nc{\bg}{B. Grzadkowski}
\nc{\BG}{Bohdan Grzadkowski}
\nc{\lsp}{\;\;\;\;\;\;\;\;}
\nc{\beq}{\begin{equation}}   \nc{\eeq}{\end{equation}}
\nc{\bea}{\begin{eqnarray}}   \nc{\eea}{\end{eqnarray}}
\nc{\baa}{\begin{array}}      \nc{\eaa}{\end{array}}
\nc{\bit}{\begin{itemize}}    \nc{\eit}{\end{itemize}}
\nc{\ben}{\begin{enumerate}}  \nc{\een}{\end{enumerate}}
\nc{\bce}{\begin{center}}     \nc{\ece}{\end{center}}
\nc{\non}{\nonumber}
\nc{\lumun}{\;{\hbox {fb}^{-1}}{\hbox {yr}^{-1}}}
\nc{\hc}{\hbox {h.c.}}
\nc{\re}{\hbox {Re}}
\nc{\im}{\hbox {Im}}
\nc{\etal}{\hbox{et al.}}
\nc{\pbarn}{\;\hbox {pb}}
\nc{\prd}[3]{{\it Phys.\ Rev.}\ {{\bf D{#1}} (#2) #3}}
\nc{\prl}[3]{{\it Phys.\ Rev.\ Lett.}\ {{\bf {#1}} (#2) #3}}
\nc{\plb}[3]{{\it Phys.\ Lett.}\ {{\bf B{#1}} (#2) #3}}
\nc{\npb}[3]{{\it Nucl.\ Phys.}\ {{\bf B{#1}} (#2) #3}}
\nc{\ptp}[3]{{\it Prog.\ Theor.\ Phys.}\ {{\bf {#1}} (#2) #3}}
\nc{\zfp}[3]{{\it Z.\ Phys.}\ {{\bf C{#1}} (#2) #3}}
\nc{\mpla}[3]{{\it Mod.\ Phys.\ Lett.}\ {{\bf A{#1}} (#2) #3}}
\nc{\rmp}[3]{{\it Rev.\ Mod.\ Phys.}\ {{\bf {#1}} (#2) #3}}
\nc{\ijmpa}[3]{{\it Int.\ J.\ of\ Mod.\ Phys.}\
               {{\bf A{#1}} (#2) #3}}
\nc{\app}[3]{{\it Acta\ Phys.\ Polon}\ {{\bf B{#1}} (#2) #3}}
\nc{\epj}[3]{{\it Eur. Phys. J.}\ {{\bf C{#1}} (#2) #3}}
\nc{\ra} {\rightarrow}
\nc{\cw}{\cos\theta_W}        \nc{\sw}{\sin\theta_W}
\nc{\ttbar}{t\bar{t}}
\nc{\bbbar}{b\bar{b}}
\nc{\tanb} {\tan \beta}
\nc{\twbdec} {t\rightarrow W^+ b}
\nc{\tbwbdec} {\bar{t} \rightarrow W^- \bar{b}}
\nc{\hprod} {e^+e^- \ra Z^\ast \ra H Z}
\nc{\epem} {e^+e^-}
\nc{\wpwm} {W^+W^-}
\nc{\tbar} {\bar{t}}
\nc{\bbar} {\bar{b}}
\nc{\wpp} {W^+}
\nc{\mt}{m_t}
\nc{\mts}{m_t^2}
\nc{\mw} {m_W}
\nc{\mws} {m_W^2}
\nc{\mz} {m_Z}
\nc{\mzs} {m_Z^2}
\nc{\mh} {m_H}
\nc{\mhs} {m_H^2}
\nc{\ma} {m_A}
\nc{\mas} {m_A^2}
\nc{\hdec}{H \ra t\bar{t}}
\nc{\ttbardec}{\ttbar \ra W^+W^-\bbbar}
\nc{\po}{\Phi_1}
\nc{\pod}{\Phi_1^\dagger}
\nc{\pht}{\Phi_2}
\nc{\phtd}{\Phi_2^\dagger}
\nc{\phtt}{{\tilde{\Phi}}_2}
\nc{\popo}{\po^\dagger\po}
\nc{\phtpt}{\pht^\dagger\pht}
\nc{\popt}{\po^\dagger\pht}
\nc{\phtpo}{\pht^\dagger\po}
\nc{\sq}{\sqrt{2}}
\nc{\nsd} {N_{SD}}
\nc{\ntt} {N_{tt}}
\nc{\vs}{\vspace{2mm}}
\nc{\sty}{\hat{S}^t_1} \nc{\pty}{\hat{P}^t_1}
\nc{\sts}{(\sty)^2}      \nc{\pts}{(\pty)^2}
\nc{\yts}{\sts+\pts}
\nc{\sby}{\hat{S}^b_1} \nc{\pby}{\hat{P}^b_1}
\nc{\sbs}{(\sby)^2}      \nc{\pbs}{(\pby)^2}
\nc{\ybs}{\sbs+\pbs}

\def\wmp{W^{\mp}}

\def\ie{{\it i.e.}}
\def\eg{{\it e.g.}}
\def\sb{s_\beta}
\def\cb{c_\beta}
\def\rts{\sqrt s}

\def\hsm{h_{\rm SM}}
\def\mhsm{m_{\hsm}}

\def\h{h}

\def\hl{h^0}
\def\hh{H^0}
\def\ha{A^0}

\def\hpm{H^{\pm}}
\def\mhl{m_{\hl}}
\def\mhh{m_{\hh}}
\def\mha{m_{\ha}}

\def\lsim{\mathrel{\raise.3ex\hbox{$<$\kern-.75em\lower1ex\hbox{$\sim$}}}}
\def\gsim{\mathrel{\raise.3ex\hbox{$>$\kern-.75em\lower1ex\hbox{$\sim$}}}}
\def\anti{\overline}

\def\fbi{~{\rm fb}^{-1}}

\def\mev{\,{\rm MeV}}
\def\gev{\,{\rm GeV}}
\def\tev{\,{\rm TeV}}

\begin{document}
%
\font\fortssbx=cmssbx10 scaled \magstep2
\medskip
\begin{flushright}
$\vcenter{
\hbox{\bf UCD-99-1} 
\hbox{\bf IFT-28-98}
\hbox{\bf hep-ph/9902308}
\hbox{February, 1999}
}$
\end{flushright}
\vspace*{2cm}
\begin{center}
{\large{\bf Finding the CP-Violating Higgs Bosons  at
{\boldmath $\epem$} Colliders}}\\ 
\rm
\vspace*{1cm}
\renewcommand{\thefootnote}{\alph{footnote})}
{\bf \BG$^1$}
\footnote{E-mail:{\tt bohdang@fuw.edu.pl}} 
{\bf John F. Gunion$^2$}
\footnote{E-mail:{\tt jfgucd@pc90.ucdavis.edu}} 
and {\bf Jan Kalinowski$^1$}
\footnote{E-mail:{\tt kalino@fuw.edu.pl}}\\

\vspace*{1.5cm}
{$^1$ \em Institute of Theoretical Physics, Warsaw University, 
Warsaw, Poland}\\
{$^2$ \em Davis Institute for High Energy Physics, 
UC Davis, CA, USA }\\

\vspace*{1.5cm}

{\bf Abstract}
\end{center}
\vspace{5mm} We discuss a general two-Higgs-doublet model with CP
violation in the Higgs sector. In general, the three neutral Higgs fields 
of the model all mix and the resulting physical Higgs bosons have no definite CP 
properties.  We derive a new sum rule relating Yukawa
and Higgs--$Z$ couplings which implies that a neutral Higgs
boson cannot escape detection at an $e^+e^-$ collider 
if it is kinematically accessible in $Z$+Higgs, $b\anti b+$Higgs
and $t\anti t+$Higgs production, irrespective of the mixing
angles and the masses of the other neutral Higgs bosons.
We also discuss modifications of the sum rules and
their phenomenological consequences in the case when the 
two-doublet Higgs sector is extended by adding one or more singlets.

\vfill
\setcounter{page}{0}
\thispagestyle{empty}
\newpage

\renewcommand{\thefootnote}{\sharp\arabic{footnote}}
\setcounter{footnote}{0}

\section{Introduction}

Despite the spectacular successes of high-energy physics (\eg.
precision tests of the Standard Model), the origins of mass and of CP
violation still remain mysteries from both the experimental and
the theoretical points of view. Models of mass generation by electroweak
symmetry breaking driven by elementary scalar dynamics predict the
existence of one or more physical Higgs bosons. The minimal model
is a one-doublet Higgs sector as employed in the Standard
Model (SM), which gives rise to fermion masses and to a single
physical CP-even Higgs scalar boson, $\hsm$. But, a Higgs boson
has yet to be observed. Regarding CP, there is only
one solid experimental signal of CP violation, namely
$K_L^0\ra\pi^+\pi^-$ decay~\cite{expcp}. The classical method for
incorporating CP violation into the SM is to make 
the Yukawa couplings of the Higgs boson to
quarks explicitly complex,  as  built into the Kobayashi-Maskawa
mixing matrix \cite{km} proposed more than two decades ago. 
However, CP violation could equally well be partially or wholly due
to other mechanisms. The possibility that 
CP violation derives largely from the Higgs sector 
itself is particularly appealing. Even 
the simple two-Higgs-doublet model (2HDM) extension of the 
one-doublet SM Higgs sector
provides a much richer framework for describing CP violation; in the 2HDM,
spontaneous and/or explicit CP violation is possible
in the scalar sector~\cite{weinberg}.

The CP-conserving (CPC) version of the 
2HDM has received considerable attention,
especially in the context of the minimal supersymmetric model (MSSM)
\cite{hhg}. It predicts~\footnote{However, with
  soft-supersymmetry $CP$-violating phases, the $\hl$, $\hh$ and $\ha$
  will mix beyond the Born approximation~\cite{cp-phases}.} 
the existence of two neutral CP-even Higgs
bosons ($\hl$ and $\hh$, with $\mhl\leq\mhh$), one neutral CP-odd
Higgs ($\ha$) and a charged Higgs pair ($\hpm$). However, in a general
2HDM with CP-violation (CPV) in the scalar sector, the three electrically
neutral Higgs fields mix and the physical mass eigenstates, $\h_i$
($i=1,2,3$), have undefined CP properties.

The absence of any $\epem\to Z \hsm$ signal in LEP1 data (where the
$Z$ is virtual) and LEP2 data (where the $Z$ is real) translates into
a lower limit on $\mhsm$ which has been increasing as higher energy
data becomes available. The latest analysis of four LEP experiments
at $\sqrt{s}$ up to 189 GeV 
implies $\mhsm$ greater than 87.5 GeV (ALEPH), 94.1 GeV (DELPHI), 95.5
GeV (L3), 94.0 GeV (OPAL) \cite{newdata}. The negative results of Higgs
boson searches at LEP can be formulated as restrictions on the
parameter space of the 2HDM and more general Higgs sector models. As has
been shown in Ref.~\cite{gghk}, the sum rules for the Higgs--$Z$ boson
couplings derived in the CP-conserving 2HDM can be generalized to the
CP-violating case 
to yield a sum rule [see Eq.~(\ref{oldsr})] that requires at least
one of the $ZZ\h_i$, $ZZ\h_j$ and $Z \h_i\h_j$ ({\it any} $i\neq j$,
$i,j=1,2,3$) couplings to be substantial in size. 
Very roughly, this implies that if there are two light Higgs bosons 
with $m_{h_i}+ m_{h_j}$, $m_{h_i}+m_Z$ and $m_{h_j}+m_Z$
all sufficiently below $\sqrt{s}$, then at least one will be observable. 
A recent analysis of LEP data shows that the 95\%
confidence level exclusion region in the $(m_{\h_i},m_{\h_j})$ plane
that results from the general sum rule is quite significant
\cite{cpvdelphi}. 

It is also appropriate to consider the
implications of the precision LEP and Tevatron electroweak data
for the general 2HDM. In the 
context of the SM, $\mhsm\leq 260\gev$
is required for $\Delta \chi^2\leq (1.64)^2$ (corresponding
to 95\% CL for a one-sided distribution)~\cite{lepewwglimit}.
In the 2HDM, any neutral Higgs boson with significant
$ZZh$ coupling ($g_{WWh}/g_{ZZh}$ is the same as in the SM) 
contributes to $\Delta \rho$ an amount given by $g_{ZZh}^2/g_{ZZ\hsm}^2$
times the contribution of a SM Higgs boson of the same mass.
In the absence of additional contributions to $\Delta\rho$, the SM
limit roughly converts to the requirement that at least one
of the neutral $h_i$ must have mass below $260\gev$ {\it and} have
substantial $ZZh_i$ coupling. However, if the Higgs bosons
of the 2HDM are not all degenerate, there can be additional
positive contributions to $\Delta\rho$ which compensate an enhanced
negative contribution to $\Delta\rho$ (by virtue of larger $m_h$)
from the diagrams involving the $ZZh$ and $WWh$ couplings.
Very roughly \cite{hhg}, substantial extra contributions arise
when there is a (neutral) $h_i$ with $|m_{h_i}-m_{\hpm}|$ 
and $g_{h_i\hpm\wmp}$ both large or when there is a neutral $h_i$ - $h_j$
pair with $|m_{h_i}-m_{h_j}|$ and $g_{h_ih_jZ}$ both large. In the MSSM,
one is protected against such situations by the natural `decoupling'
limit of the model. In the general 2HDM, significant extra positive
$\Delta\rho$ contributions are possible in a general scan over model
parameters.  Thus, constraints from the precision data
are complicated and will not be directly implemented here.

In this paper, we consider the 2HDM in the
context of higher energy $e^+e^-$ linear colliders ($\rts\sim 350-1600\gev$).
The general question we wish to address is whether we
are guaranteed to see any neutral Higgs boson that is light.
Two scenarios give cause for concern. 
\bit
\item First, the precision
electroweak suggestion of a light $h_i$ with significant $ZZh_i$ coupling
could prove correct, in which case 
the $h_i$ will be seen in $\epem\to Z^*\to Zh_i$
Higgs-strahlung production. However, it could happen that
there are actually two light Higgs bosons. We denote the second by $h_j$. 
There are then two possibilities allowed by
the above-mentioned sum rule [Eq.~(\ref{oldsr})]. (a) If the $h_i$
observed in $Zh_i$ does not have full strength $ZZh_i$ coupling then 
either the $Zh_ih_j$ or $ZZh_j$ coupling (or both)
must be substantial and $h_j$ will be observable in the $h_ih_j$ or $Zh_j$
final state (or both) provided 
$m_{h_i}+m_{h_j}<\sqrt s-\Delta$ and $m_{h_j}+m_Z<\sqrt s-\Delta^\prime$,
where $\Delta$ and $\Delta^\prime$ generically
represent the subtractions from the absolute kinematic limits due to
backgrounds, efficiencies and finite luminosity. (b) If the $h_i$
has full strength $ZZh_i$ coupling, then the sum rule guarantees
that the $Zh_ih_j$ and $Zh_j$ couplings vanish and, therefore,
the $h_j$ will not be discovered via Higgs-strahlung ($Zh_j$) or pair
($h_ih_j$) production. (Note that the above conclusions hold
regardless of the mixing structure of the neutral Higgs boson sector.)
It is case (b) that causes concern.
\item
A second, and even worse scenario, is the following.
It could happen that there is only one light $h_i$ but model parameters
conspire so that it has a $ZZh_i$ coupling
that is too weak for its detection in Higgs-strahlung production
while at the same time precision electroweak constraints are satisfied.
\eit
The primary result of the present paper is the derivation
of new sum rules that relate the Yukawa and Higgs--$Z$
couplings of the 2HDM [see Eq.~(\ref{yuksr})] in such a way as to guarantee
that any $h_i$ that is sufficiently light ($m_{h_i}+2m_t< \rts-\Delta$)
will be observable regardless of the mixing structure of the neutral Higgs
boson sector and independent of the masses of the other Higgs bosons.
Very roughly, this new sum rule implies that if the Higgs-strahlung
cross section for $h_i$ is small because of small $ZZh_i$ coupling,
then the cross section for either $b\anti b h_i$ or $t\anti t h_i$
(dominated by Higgs radiation from the final state fermions) will
be large enough to be detected. 

We shall also discuss the extension of
these sum rules to the two-doublet plus one-singlet CP-violating model.
We find that there is no guarantee that a single light Higgs boson will
be observable.  However, the extended sum rules do imply
that if there are three light (as defined above) Higgs bosons, then 
at least one will be observable via
production in association with $b\anti b$ or $t\anti t$.

Before proceeding, it should be emphasized that
our results make no assumption as
to the nature of the model at energies above the Higgs boson masses.
As shown in Ref.~\cite{espgun}, demanding perturbativity for all
couplings up to a scale of order the Planck mass places strong
constraints on the spectrum of those Higgs bosons that have substantial
$ZZ$ coupling. These constraints are such that the next generation
of $\epem$ collider would be able to see $Zh$ production for at
least one Higgs boson or collection of Higgs bosons.
Our focus here is on results that apply
purely as a result of the structure of the low-energy Higgs sector model. 

The paper is organized as follows. In Section 2, we outline how CP
violation arises in the 2HDM and give the general forms of
the $ZZ$-Higgs, $Z$-Higgs-Higgs, and Higgs Yukawa couplings
in terms of the matrix specifying the mixing of the neutral Higgs bosons.
In Section 3, we present the crucial sum rules for these couplings.
In Section 4, we specify the existing experimental constraints
that we require be satisfied as we scan over Higgs masses
and mixing parameters. Numerical results for 
$Zh_1h_2$, $b\anti b h_1$ and $t\anti t h_1$ cross sections
resulting from the scan over 2HDM parameter space
are presented and discussed in Section 5. In Section 6, we extend
the sum rules to the case of the two-doublet plus one-singlet Higgs
sector and outline implications. Concluding remarks are given in Section 7.
The Appendix presents the detailed cross section formula 
for the $\epem\to f\anti f h_i$ process allowing for Higgs boson mixing
and CP violation.

\section{The two-Higgs-doublet model with CP violation}

The 2HDM of electroweak interactions 
 contains two SU(2) Higgs doublets denoted by 
$\Phi_1=(\phi_1^+,\phi_1^0)$ and $\Phi_2=(\phi_2^+,\phi_2^0)$ and is
defined by Yukawa couplings and the Higgs potential. 
The most general renormalizable 
scalar potential for the model has the following form:
\begin{eqnarray}
V(\po,\pht)&=& V_{symm}(\po,\pht)+ V_{soft}(\po,\pht)+
 V_{hard}(\po,\pht) \\
\label{potential}
 V_{symm}(\po,\pht)&=&-\mu_1^2\popo-\mu_2^2\phtpt \nonumber  \\ 
&&+\lambda_1(\popo)^2+\lambda_2(\phtpt)^2+
\lambda_3(\popo)(\phtpt)\nonumber \\
&&+\lambda_4|\popt|^2+\frac{1}{2}
\left[\lambda_5(\popt)^2+\hc\right] \nonumber \\
V_{soft}(\po,\pht)&=&-\mu_{12}^2\popt+\hc  \nonumber \\ 
V_{hard}(\po,\pht)&=& \frac{1}{2}\lambda_6(\popo)(\popt)+
\frac{1}{2}\lambda_7(\phtpt)(\popt)+\hc \nonumber 
\end{eqnarray}
If both of the two Higgs boson doublets couple to
up- or to down-type quarks (or to both types), 
flavor changing neutral currents (FCNC) are generated at
tree level. To avoid FCNC, it is customary to impose a discrete $Z_2$
symmetry under which
\begin{equation}
\pht\ra-\pht,\;\;\;\;\; {u_i}_R\ra-{u_i}_R
\label{dsym}
\end{equation}
and the other fields are unchanged. Then, $\Phi_2$ couples only
to up-type quarks and $\Phi_1$ couples only to down-type quarks
and leptons. The resulting
invariant fermion-Higgs Yukawa interactions can be written in the form
\begin{equation}
{\cal L}_Y=-(\bar{u}_i,\bar{d}_i)_L \Gamma_u^{ij} \phtt {u_j}_R 
           -(\bar{u}_i,\bar{d}_i)_L \Gamma_d^{ij} \po  {d_j}_R
           -(\bar{\nu}_i,\bar{e}_i)_L \Gamma_e^{ij} \po  {e_j}_R + \hc,
\label{yukcoupl}
\end{equation}
where $i,j$ are generation indices and $\phtt$ is defined as 
$i\sigma_2 \pht^\ast$.
Only the first term $V_{symm}(\po,\pht)$ in Eq.~(\ref{potential}) 
is symmetric under $Z_2$. However, if the $Z_2$ symmetry is broken
only softly (that is by operators of dimension 3 and less) 
then renormalizability is preserved~\cite{sym} and FCNC effects remain 
small. The unique soft-breaking term is that appearing 
in $V_{soft}(\po,\pht)$. The dimension 4 
terms contained in $V_{hard}(\po,\pht)$ break
the $Z_2$ symmetry in a hard way and therefore cannot be 
accepted.~\footnote{If $V_{hard}$ is present,
there is no argument for dropping the FCNC Yukawa terms which are 
also of dimension~4.}    

The 2HDM Higgs sector can exhibit either explicit or spontaneous CP violation.
CP violation is explicit if there is no choice of phases
such that all the potential parameters are real. CP violation
is said to be spontaneous if the potential minimum is
such that one of the two vacuum expectation values is complex,
even though all the potential parameters can be chosen to be real.
If only $V_{symm}$ is present then neither explicit
nor spontaneous CP violation can be present in the Higgs sector \cite{buras}.
In fact, when FCNC are suppressed by 
imposing {\it exact} $Z_2$ symmetry, one must
introduce a third Higgs doublet in order to allow for
CP violation in the Higgs sector.
However,  both explicit and spontaneous 
CP violation in the 2HDM become possible 
even if the $Z_2$ symmetry is only broken softly.
The CP violation will be explicit in $V_{symm}+V_{soft}$ if
$\im({\mu_{12}^\ast}^4\lambda_5)\neq 0$. 
When $\im({\mu_{12}^\ast}^4\lambda_5)= 0$, spontaneous CP violation
can arise as follows. Without loss of generality, the phase of $\po$
can be chosen such that its vacuum expectation value is real and positive,
$<\po>=v_1/\sqrt{2}$ (with $v_1>0$), 
and the phase of $\pht$ such that the $\lambda_5$ coupling is real
and positive. Then,
the second Higgs doublet will have a complex vacuum expectation value, 
$<\pht>=v_2e^{i\theta}/\sqrt{2}$ ($v_2>0$ by convention),~\footnote{
In this normalization $v\equiv \sqrt{v_1^2+v_2^2}=2m_W/g=246\,\mbox{GeV}$.}
provided
\beq
\left|\frac{\mu_{12}^2}{2\lambda_5v_1v_2}\right|<1,
\eeq
since, then, the minimum of the potential occurs for~\cite{leephase}
\beq
\cos\theta=\frac{\mu_{12}^2}{2\lambda_5v_1v_2}.
\label{cpphase}
\eeq 
Therefore, the 2HDM with Higgs potential given by $V_{soft}+V_{symm}$
is a very attractive and simple model in which to explore
the implications of CP violation in the Higgs sector. 

After SU(2)$\times$U(1) gauge symmetry breaking, one combination of neutral
Higgs fields, $\sqrt2(\cb\mbox{Im}\phi_1^0+ \sb\mbox{Im}\phi_2^0)$,
becomes a would-be Goldstone boson which is absorbed in giving 
mass to the $Z$ gauge boson.
(Here, we use the notation $\sb\equiv\sin\beta$, $\cb\equiv\cos\beta$,
where $\tanb=v_2/v_1$.)
The same mixing angle, $\beta$, also diagonalizes 
the mass matrix in the charged Higgs sector.  
If either explicit or spontaneous CP violation is present,
the remaining three neutral degrees of freedom, 
\begin{equation}
(\varphi_1,\varphi_2,\varphi_3)\equiv
\sqrt 2(\mbox{Re}\phi_1^0, \, \mbox{Re}\phi_2^0, \,
 s_\beta\mbox{Im}\phi_1^0-c_\beta\mbox{Im}\phi_2^0) 
\end{equation} 
are not mass eigenstates. The physical neutral Higgs bosons $h_i$
($i=1,2,3$) are obtained by an orthogonal transformation, $h=R
\varphi$, where the rotation matrix is given in terms of three Euler
angles ($\alpha_1, \alpha_2,\alpha_3$) by
\begin{eqnarray} 
R=\left(\baa{ccc}
  c_1     &  -s_1c_2          &     s_1s_2  \\
  s_1c_3  & c_1c_2c_3-s_2s_3  &  -c_1s_2c_3-c_2s_3\\
  s_1s_3 & c_1c_2s_3+s_2c_3 & -c_1s_2s_3+c_2c_3 \eaa\right),
\label{mixing}
\end{eqnarray}
where $s_i\equiv\sin\alpha_i$ and $c_i\equiv\cos\alpha_i$.
Without loss of generality, we assume $m_{h_1}\le m_{h_2} \le m_{h_3}$. 

As a result of the mixing between real and imaginary parts of neutral
Higgs fields, the Yukawa interactions of the $h_i$ mass-eigenstates are not
invariant under CP. They are given by:
\begin{equation} 
{\cal L}=h_i\bar{f}(S^f_i+iP^f_i\gamma_5)f \label{coupl} 
\end{equation}
where the scalar ($S^f_i$) and pseudoscalar ($P^f_i$) couplings are
functions of the mixing angles. For up-type quarks we have 
\begin{equation}
S^u_i=-\frac{m_u}{v s_\beta}R_{i2},\;\;\;\;\;
P^u_i=-\frac{m_u}{v s_\beta}c_\beta R_{i3}, 
\label{absu}
\end{equation}
and for down-type quarks one finds
\begin{equation}
S^d_i=-\frac{m_d}{v c_\beta}R_{i1},\;\;\;\;\;
P^d_i=-\frac{m_d}{v c_\beta}s_\beta R_{i3}\,,
\label{absd}
\end{equation}
and similarly for charged leptons. 
For large $\tan\beta$, the couplings to down-type fermions are 
typically enhanced over the couplings to up-type fermions.

In the following analysis we will also need  
the couplings of neutral Higgs and $Z$ bosons; they are given by
\begin{eqnarray}
g_{ZZh_i}& \equiv & \frac{g m_Z}{c_W} C_i= 
\frac{g m_Z}{c_W} (s_{\beta} R_{i2}+c_{\beta}R_{i1})
\label{zzhcoup}
\\
g_{Zh_ih_j}& \equiv & \frac{g}{2c_W} C_{ij} =
\frac{g}{2c_W}(w_i R_{j3}-w_j R_{i3})
\label{zhhcoup}
\\
g_{ZZh_ih_j}&\equiv& \frac{g^2}{2c_W^2} X_{ij}=
\frac{g^2}{2c_W^2}\sum_{k=1}^3 R_{ik}R_{jk}
\label{zzhhcoup}
\end{eqnarray}
where $w_i=s_{\beta}R_{i1}-c_{\beta}R_{i2}$,   
$c_W=\cos\theta_W$, $g$
is the SU(2) gauge coupling constant and $m_Z$ denotes the $Z$-boson mass.
In the case of the 2HDM, $X_{ij}=\delta_{ij}$
by virtue of the orthogonality of $R$ and its $3\times 3$ dimensionality;
in particular, the $ZZh_ih_j$ coupling is not
suppressed by mixing angles.

The CP-conserving limit can be obtained as a special case: $\alpha_2=
\alpha_3=0$. Then, if we take $\alpha_1=\pi/2-\alpha$, $\alpha$ is the 
conventional mixing angle 
that diagonalizes the mass-squared matrix for $\sqrt 2\mbox{Re}\phi_1^0$ and
$\sqrt 2\mbox{Re}\phi_2^0$.
The resulting  mass eigenstates are $h_1=-\hl$ $h_2=\hh$ and 
$\sqrt 2(s_\beta\mbox{Im}\phi_1^0-c_\beta\mbox{Im}\phi_2^0)=-\ha$, 
where $\hl$, $\hh$ ($\ha$) are the CP-even (CP-odd) Higgs bosons
defined earlier for the CPC 2HDM.

\section{Sum rules for the Higgs boson couplings}

As discussed earlier, we wish to determine whether or not
the additional freedom in Higgs boson couplings in the general
CP-violating 2HDM 
(by tuning the mixing angles one can suppress certain couplings) 
is sufficient to jeapordize our ability to find light neutral Higgs bosons.
We will show that the unitarity of $R_{ij}$ 
implies a number of interesting sum rules for the Higgs  
couplings which prevent the hiding of any neutral Higgs boson
that is sufficiently light to be kinematically accessible
(a) in Higgs-strahlung {\it and} Higgs pair production, or (b) Higgs-strahlung
{\it and} $b\anti b$+Higgs {\it and} $t\anti t$+Higgs. 
\begin{description}
\item{a)}
Let us first recall the sum rule for Higgs--$Z$ 
couplings that requires at least one of the $ZZ\h_i$, $ZZ\h_j$ and $Z
\h_i\h_j$ ({\it any} $i\neq j$, $i,j=1,2,3$) couplings to be
substantial in size \cite{gghk}, namely
\begin{equation}
  \label{oldsr}
C_{i}^2+C_{j}^2+C_{ij}^2=1  
\end{equation}
where $i\neq j$ are any two of the three possible
indices.~\footnote{Another interesting sum rule reads $C_{ij}^2=C_k^2$
for $(i,j,k)$ being any permutation of (1,2,3).} 
The power of Eq.~(\ref{oldsr})
with $i,j=1,2$ for LEP physics derives from two facts: it
involves only two of the neutral Higgs bosons; and the
experimental upper limit on any one $C_i^2$ derived from $\epem\to
Z\h_i$ data is very strong --- $C_i^2\lsim 0.1$ for $m_{\h_i}\lsim 70$ GeV.
Thus, if $\h_1$ and $\h_2$ are both below about $70\gev$ in mass, then
Eq.~(\ref{oldsr}) requires that $C_{12}^2\sim 1$, whereas for such
masses the limits on $\epem\to \h_1\h_2$ from LEP2 data require
$C_{12}^2\ll 1$. As a result, there cannot be two light Higgs bosons
even in the general CP-violating case; the excluded region in the
$(m_{\h_1},m_{\h_2})$ plane that results from a recent analysis by the
DELPHI Collaboration can be found in Ref.~\cite{cpvdelphi}.

At a higher energy $\epem$ collider, Eq.~(\ref{oldsr}) will have
many possible applications. If no Higgs boson is discovered
in Higgs-strahlung or Higgs pair production, 
Eq.~(\ref{oldsr}) will imply that at least one of $m_{h_i}+m_{h_j}$, 
$m_{h_i}+\mz$ and $m_{h_j}+\mz$ must be $ > \sqrt s-\Delta$
for any choice of $i$ and $j$. However, as noted earlier, this does not
preclude the possibility that there is a light $h_i$
with $m_{h_i}+\mz<\sqrt s-\Delta$ but with small $ZZh_i$
coupling. More likely, the precision electroweak
suggestion will turn out to be correct and at the $\epem$ collider
we will find at least
one Higgs boson in $\epem\to Zh_i$ production (note that
$h_i$ need not be the lightest neutral Higgs boson) and 
measure its $C_i$ with good accuracy.
If the observed $h_i$ has $C_i\sim 1$, then Eq.~(\ref{oldsr})
implies that any other $h_j$ must have small $ZZh_j$ and $Zh_ih_j$
couplings and will not be observable in Higgs-strahlung or Higgs pair
production (in association with the observed $h_i$). If the measured
$C_i$ is substantially smaller than 1, then Eq.~(\ref{oldsr}) implies
that either $\epem\to h_ih_j$ or
$\epem \to Zh_j$ would have a substantial rate for any
sufficiently light $h_j$ ($j\neq i$). If a second $h_j$
has not been detected, we would then conclude that
$m_{h_j}>{\rm min}[\sqrt s- m_{h_i}-\Delta,
\sqrt s-m_Z-\Delta^\prime]$ for the other two $j\neq i$ neutral Higgs bosons.

\item{b)}
If even one of the three processes, $Zh_1$, $Zh_2$ (Higgs-strahlung)
and $h_1h_2$ (pair production), is beyond the collider's kinematical
reach, the sum rule in Eq.~(\ref{oldsr}) is not sufficient to
guarantee $h_1$ or $h_2$ discovery. For example, suppose that
$h_1h_2$ production is not kinematically allowed. Eq.~(\ref{oldsr})
can be satisfied by taking $C_{12}\sim 1$ and $C_{1,2} \sim 0$.
For these choices, $Zh_1$ and $Zh_2$ production would be suppressed
and unobservable (even if kinematically allowed)
because of small $C_1$ and $C_2$, respectively. 
However, we find that the
Yukawa and $ZZ$ couplings of any one Higgs boson
also obey sum rules which require that at
least one of these couplings has to be sizable; \ie\ if $C_i\sim 0$
at least one $h_i$ Yukawa coupling must be large. Thus, if an $h_i$
is sufficiently light, its detection will be possible, irrespective of 
the neutral Higgs sector mixing.

To derive the relevant sum rules, it is convenient to
introduce rescaled couplings
\begin{eqnarray}
    \label{rescal}
\hat{S}^f_i\equiv \frac{S^f_i v}{m_f}\,,~~~~~~~
\hat{P}^f_i\equiv \frac{P^f_i v}{m_f}\,,
\end{eqnarray}
where $f=t,b$.~\footnote{For obvious reasons we consider the third
generation of quarks. Similar expressions hold for for lighter
generations.} Using Eqs.~(\ref{absu}) and (\ref{absd}), one finds:
\begin{eqnarray}
  \label{yuk}
(\hat{S}^t_i)^2 + (\hat{P}^t_i)^2 
&=&\left(\frac{\cos\beta}{\sin\beta}\right)^2 \left[
R^2_{i3}+R^2_{i2}/\cos^2\beta
\right]\,;
\nonumber \\
(\hat{S}^b_i)^2 + (\hat{P}^b_i)^2 
&=&\left(\frac{\sin\beta}{\cos\beta}\right)^2 \left[
R^2_{i3}+R^2_{i1}/\sin^2\beta
\right]\,.
\end{eqnarray}
Using the unitarity of $R_{ij}$, these can be written as:
\begin{eqnarray}
  \label{yuksr}
(\hat{S}^t_i)^2 + (\hat{P}^t_i)^2 
&=&\left(\frac{\cos\beta}{\sin\beta}\right)^2 \left[
1+\frac{C_i}{\cos^2\beta}(2 \hat{S}^b_i \cos^2\beta+ C_i)
\right]\,;
\nonumber \\
(\hat{S}^b_i)^2 + (\hat{P}^b_i)^2 
&=&\left(\frac{\sin\beta}{\cos\beta}\right)^2 \left[
1+\frac{C_i}{\sin^2\beta}(2 \hat{S}^t_i \sin^2\beta+ C_i)
\right]\,.
\end{eqnarray}
From Eq.~(\ref{yuksr}), we see that if a light Higgs boson $h_i$ has
suppressed coupling to $ZZ$, $C_i\to 0$, then $(\hat S_i)^2+(\hat P_i)^2$
for the top and bottom quark rescaled couplings behaves as $\cot^2\beta$
and $\tan^2\beta$, respectively. If $C_i=\pm 1$, \ie\
full strength $ZZh_i$ coupling, one finds that $(\hat S_i)^2+(\hat P_i)^2\to1 $,
for both the top and the bottom quark couplings,
in the limit of either very large or very small $\tanb$.
More generally, combining the two sum rules, 
as written in Eq.~(\ref{yuk}), and using unitarity again, we find
\begin{eqnarray}
  \label{finalsr}
  \sin^2\beta [(\hat{S}^t_i)^2 + (\hat{P}^t_i)^2]
+ \cos^2\beta [(\hat{S}^b_i)^2 + (\hat{P}^b_i)^2]=1  
\end{eqnarray}
independently of $C_i$. Eq.~(\ref{finalsr})
implies that the Yukawa couplings to top and bottom
quarks cannot be simultaneously suppressed. As the earlier examples show,
the relative weighting is a sensitive function of both $\tan\beta$ and $C_i$.
In some sense, the most pessimistic case for measuring
the Yukawa couplings is $|C_i|=1$ in that it
forbids significant enhancement for 
either the top or the bottom Yukawa couplings ---
both are SM-like in the limit of large or small $\tanb$. 
Still, Eq.~(\ref{finalsr}) guarantees 
that, with sufficient integrated luminosity,
determination of at least one of the two Yukawa couplings
will be possible for any $h_i$ kinematically 
accessible in $t\anti t h_i$ (as well as $b\anti bh_i$) production.

\end{description}

The above makes it apparent that the complete
Higgs hunting strategy at $e^+e^-$ colliders, and at hadron colliders
as well, should include not only the Higgs-strahlung process
and Higgs pair production but also the Yukawa processes~\footnote{The
importance of the Yukawa processes in the context of a CP conserving
2HDM for large $\tan\beta$ has been stressed in the past many times
\cite{oldyukawa,dkz}.} with Higgs radiation off top and
bottom~\footnote{Looking for radiation off the tau leptons in the case
of large $\tan\beta$ may also help.} quarks. Details of this
strategy at a future $e^+e^-$ collider are discussed in Section 5. 

For definiteness, in what follows we will consider the high luminosity 
option that has been examined in the context of the
TESLA collider design, for which one expects $L=500$ fb$^{-1}$y$^{-1}$ 
at $\sqrt{s}=500$ GeV~\cite{tesla}. 
Since in the numerical analysis we will include constraints on the 
model parameters  
that result from the current experimental limits, we first briefly
discuss the experimental data that will be taken into account.

\section{Experimental constraints}

For given Higgs boson masses, we must consider all non-redundant 
values of the mixing angles $\alpha_i$.
Existing data already exclude certain
configurations of masses and angles, see \eg\ \cite{gghk,cpvdelphi}.
We will follow the method  used in
Ref.~\cite{gghk}, with updated experimental input. 
The constraints that we impose on the mixing angles are as follows:
\begin{itemize}

\item The $C_i^2$ are restricted by non-observation of
Higgs-strahlung events at LEP1 and LEP2. We take the limits
presented in  Fig.~16 of
Ref.~\cite{opal_limit} for  the case when no
$b$-tagging  has been used. By doing this, we avoid 
potential problems concerning the dependence of the 
Higgs-$\bbbar$ and Higgs-$\tau^+\tau^-$ branching ratios on the mixing
angles.~\footnote{We thank F.~Richard for discussions on this point.}

\item The contribution to the total $Z$-width from $Z \ra Z^* h_i \ra f
\bar{f} h_i$ (summed over $i=1,2,3$) and $Z \ra h_ih_j$ (summed
over $i,j=1,2,3: i>j$) is required to be below $7.1 \mev$; see
Ref.~\cite{opal_width}.

\item For any given values of $(m_1,m_2)$ and the $\alpha_i$, we
calculate the number of expected events  
in the processes $\epem \ra h_1 h_2 \to \bbbar \bbbar+ \bbbar \tau^+\tau^-$ 
at the LEP2 energies
$\sqrt{s}=133$, 161, 170, 172, 183 $\gev$ using the corresponding 
integrated luminosities 
$L=5.2$, 10.0, 1.0, 9.4, 54 $\pbarn^{-1}$, assuming
efficiency $\epsilon=.52$ in the individual $b\bar{b}b\bar{b}$ and
$b\bar{b}\tau^+\tau^-$ channels.  Our calculations take into account the
mixing-angle dependence of the Higgs-boson branching ratios to $\bbbar$ or
$\tau^+ \tau^-$.~\footnote{In the previous
analysis~\cite{gghk} the SM branching ratios for the Higgs boson
decays  were used.  We find, however, that our final results 
for cross sections are nearly insensitive to this modification.} 
If  the probability of observing 
zero events (after summing the rates for all energies) is below 5\%, 
the set of  masses and mixing angles is assumed to be excluded.
\end{itemize}

\section{\boldmath Higgs boson production in $e^+e^-$ colliders}

As we argued above, in $e^+e^-$ collisions production of light neutral Higgs
boson(s) can proceed via three important mechanisms: (a)
bremsstrahlung off the $Z$ boson, $e^+e^-\ra Zh_1$, (b) Higgs pair
production, $e^+e^-\ra h_1h_2$, and (c) the Yukawa processes with
Higgs radiation off a heavy fermion line in the final state,
$e^+e^-\ra f\bar{f}h_1$. The Yukawa processes are particularly
important if (a) is dynamically suppressed by the mixing and (b) is
kinematically forbidden.

\begin{figure}[h]
\leavevmode
\epsfxsize=5.5in
\centerline{\epsffile{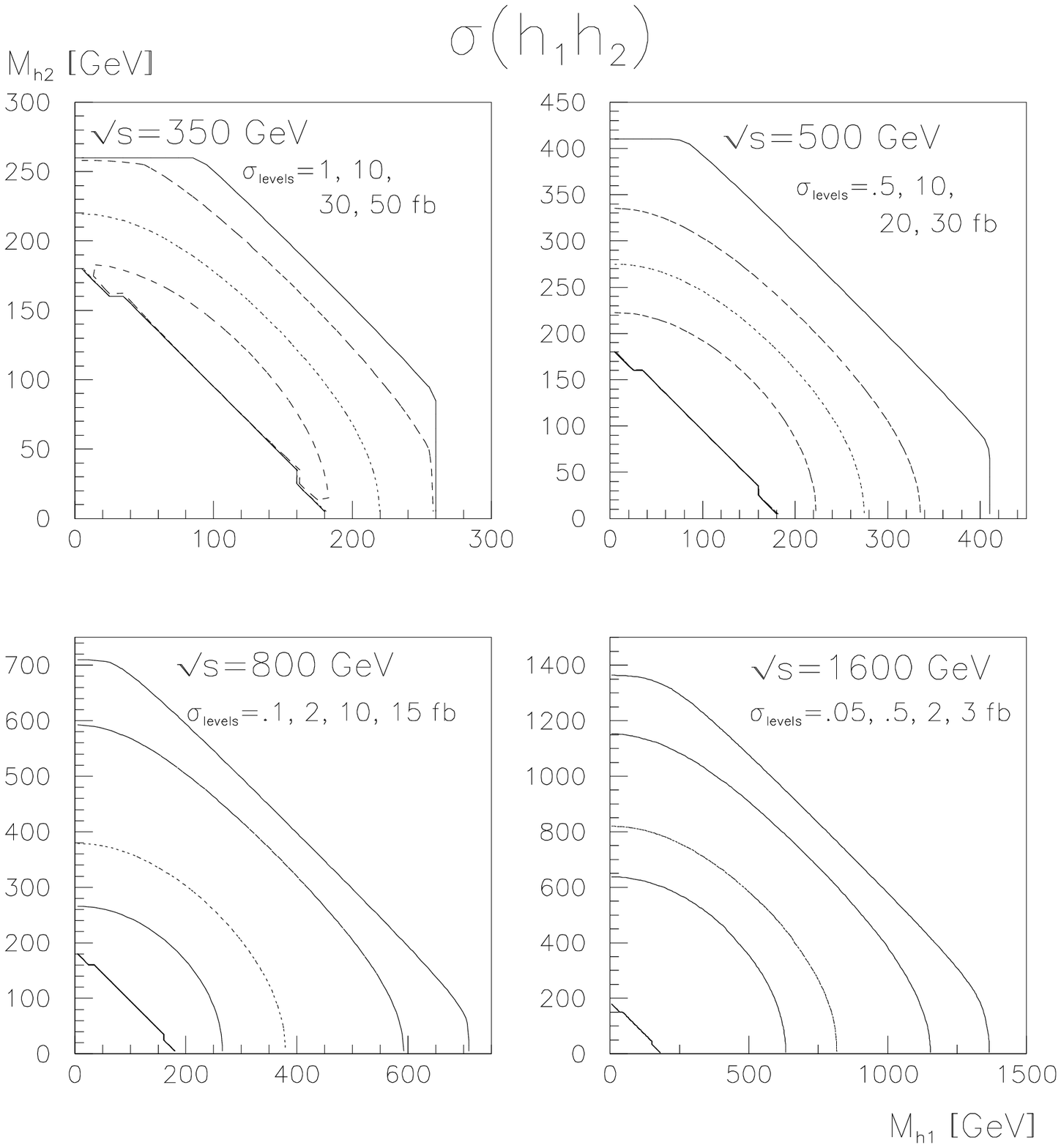}}
\medskip
\caption{Contour lines for ${\rm min}[\sigma(e^+e^-\rightarrow h_1 h_2)]$ 
as functions of Higgs boson masses for the indicated $\protect\sqrt{s}$ 
values. In scanning over mixing angles $\alpha_i$, we respect the
experimental constraints listed in Section 4, and we assume that at
any $\protect\sqrt{s}$ the number of $e^+e^- \rightarrow Zh_1$ or $Zh_2$
events is less than 50 for total luminosity $L=500 \fbi$. The
contour lines are plotted for $\tanb=0.5$; the plots are
virtually unchanged for larger values of $\tanb$. The contour lines 
overlap in the inner corner of each plot as a result of
excluding mass choices inconsistent with experimental 
constraints from LEP2 data.}
\label{h1h2_cpodd_contour}
\end{figure}

In order to treat the three processes on the same footing, we will
discuss the production of $h_1$ in association with heavy fermions:
\begin{equation}
  \label{xsec}
e^+e^- \ra f\bar{f} h_1 \,.
\end{equation}  
Feynman diagrams for processes (a) and (b) contribute to this final state 
when $Z\to f\anti f$ and $h_2\to f\anti f$, respectively. 
If $|C_1|$ is not too near 1,
Eqs.~(\ref{absu},\ref{absd}) imply that radiation diagrams (c) are enhanced
when the Higgs boson is radiated off top quarks for small $\tan\beta$
and off bottom quarks or $\tau$ leptons for large values of
$\tan\beta$.

Since all fermion and Higgs boson masses in the final state must be
kept nonzero, the formulae for the cross section are quite involved.
For the CPC case, they can be read off from Ref.~\cite{dkz}. In the CPV
case, they are more complicated due to mixing of all neutral Higgs
bosons. Therefore, for completeness, we will present the formula for
the cross section. Let $Q_f$ denote the electric charge, $N_c$ the
number of colors, $a_f$ and $v_f$ the axial and vector $Z$ charges of
the fermion $f$ normalized as
\begin{equation}
a_f=\frac{2I^f_L}{4 s_W c_W}\,, \;\;\;\; 
v_f=\frac{2I^f_L-4 Q_f s_W^2}{4 s_W c_W}\,,
\end{equation}
with $I^f_L=\pm 1/2$ being the weak isospin of the left-handed fermions.
The total cross section for the process (\ref{xsec}) can be 
written as follows:
\begin{eqnarray}
 \sigma&=&\int dx_1dx_2N_c \frac{\sigma_0}{4 \pi}
   \Biggl\{\biggl[q_e^2 q_f^2+2\frac{q_e q_f v_e v_f (1-z)}
   {(1-z)^2 +z \gamma_z}
    +\frac{(v_e^2+a_e^2) (v_f^2
      +a_f^2)}{(1-z)^2+z \gamma_z}\biggr] (G_1+F_1) \non \\
   &\;&\phantom{dx_1 dx_2 N_c \frac{\sigma_0}{4 \pi}}
    +\frac{a_e^2+v_e^2}{(1-z)^2+z \gamma_z}
      \biggl[a_f^2 (G_2+F_2+G_3+G_4+G_5+G_6)\non \\
    &\;& \phantom{dx_1 dx_2 N_c \frac{\sigma_0}{4 \pi}}
    +v_f^2 (G_4+G_6)
    + \frac{1}{16 s_W^2 c_W^2} (G_7+F_3)
       +\frac{a_f}{4s_W c_W} (F_4 + G_8)\biggr]\non \\
   &\;&\phantom{dx_1 dx_2 N_c \frac{\sigma_0}{4 \pi}}
    + \frac{q_f q_e v_e v_f (1-z)}{(1-z)^2+z\gamma_z} G_6\Biggr\},
\label{diffcross}
\end{eqnarray}
where $\sigma_0=4\pi\alpha^2/3s$ is the standard normalization cross
section. Here, $\sqrt{s}$ is the total c.m. energy,
$x_{1,2}=2E_{f,\bar{f}}/\sqrt{s}$ are the reduced energies of fermions
in the final state and $z=m_Z^2/s$, $\gamma_z=\Gamma_Z^2/s$ are the
reduced mass and width of the $Z$ boson, respectively. The functions
$G_i$ and $F_i$ are given in the Appendix: $G_{1,2}$ and $F_{1,2}$
arise from squaring graphs where $h_1$ is radiated
from the fermion; $G_{3,4}$ arise from squaring $Z\to Zh_1$ graphs; $G_{5,6}$
arise from interference between fermion-radiation and $Zh_1$ graphs;
the remaining $G$'s and $F$'s involve Higgs pair production graphs
and their interference with fermion-radiation and $Zh_1$ graphs.

\begin{figure}[h]
\leavevmode
\epsfxsize=5.5in
\centerline{\epsffile{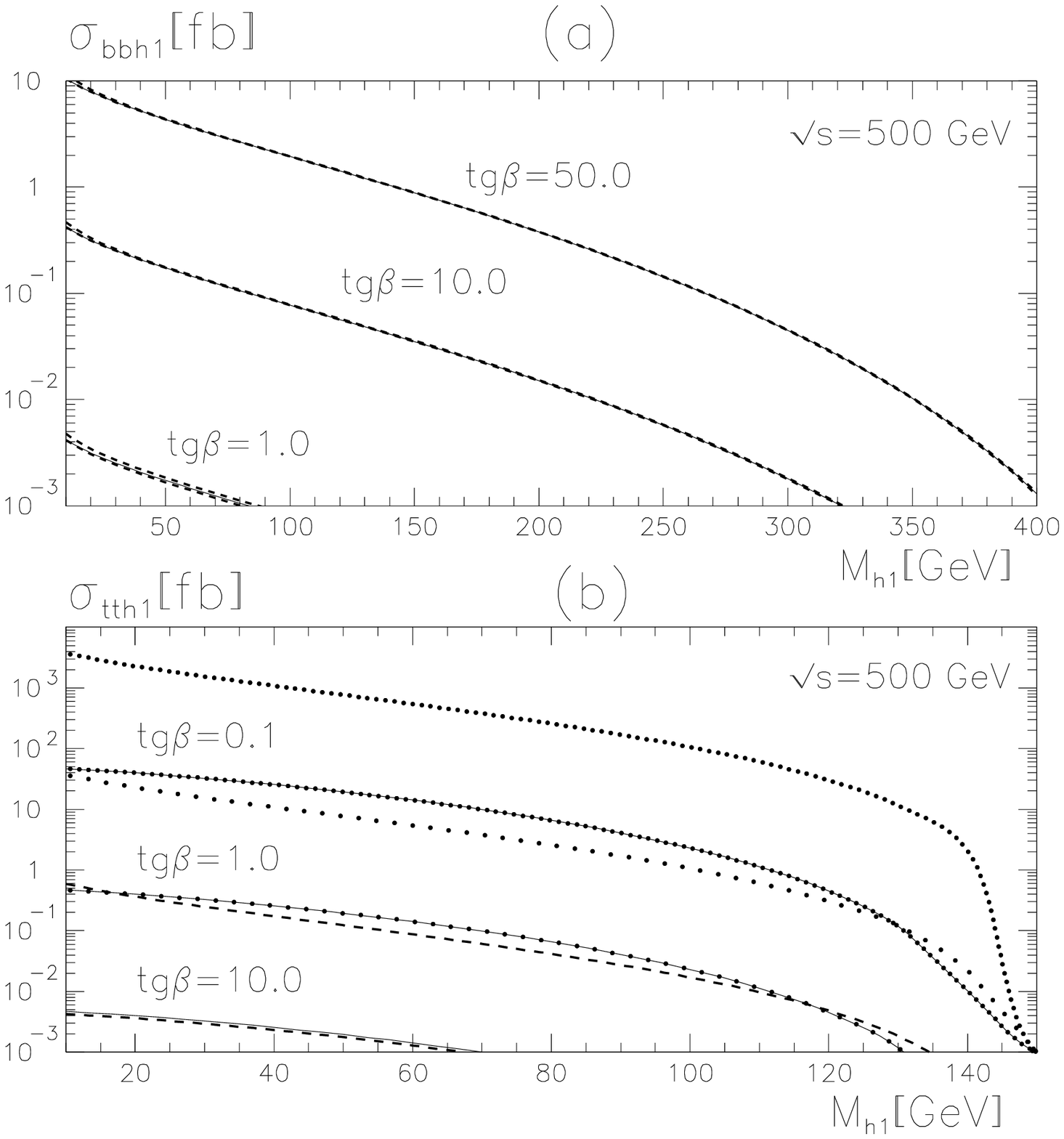}}
\medskip
\caption{
The minimal and maximal values (after
requiring fewer than 50 $Zh_1$ events for $L=500\fbi$) 
of the cross sections for $e^+e^- \rightarrow 
\bbbar h_1$ (a) and $e^+e^-\rightarrow \ttbar h_1$ (b) are plotted 
for $\protect\sqrt{s}=500$ GeV. For a given value of $\tanb$,
the same type of line (closely spaced dots for $\tanb=0.1$, 
widely spaced dots for $\tanb=1$, dashes for $\tanb=10$)
is used for the minimal and maximal values of the cross sections.
Solid lines denote cross sections
for $e^+e^-\rightarrow f\bar{f}\ha$ in the CP-conserving limit of the
general 2HDM with $\mha=m_{h_1}$. In the case of $\bbbar h_1$, the
minimal and maximal values of the cross sections are almost the same
and are almost hidden by the $\ha$ curves with the same $\tanb$ value. 
In the case of $\ttbar$, 
the minimal cross section curves are almost hidden by the $\ha$ curves
with the same $\tanb$ value.
Masses of the remaining Higgs bosons are assumed to be $1000\gev$.}
\label{fig_limits}
\end{figure}

If the coupling of the $h_1$ to the $Z$ boson is not
dynamically suppressed, \ie\ $C_1$ is substantial,
then the Higgs-strahlung process,
$e^+e^-\rightarrow Zh_1$, will be sufficient to find it. In the
opposite case, our focus in this paper, one has
to consider the other processes (b) and/or (c), for which the sum rules
(\ref{oldsr}) and (\ref{yuksr}) will imply that
the neutral Higgs boson(s), if kinematically accessible, will be
produced at a comfortably high rate at a high luminosity future linear $e^+e^-$
collider. Below we will consider two situations: (i) two light Higgs
bosons, and (ii) one light Higgs boson. 
\begin{description}
\item{(i)}
$m_{h_1}+m_{h_2},m_{h_1}+\mz,m_{h_2}+\mz<\sqrt{s}$:

If the Higgs-strahlung
processes are suppressed by mixing angles, $C_1,\, C_2\ll1$, then from
Eq.~(\ref{oldsr}) it follows that Higgs pair production is at
full strength, $C_{12}\sim 1$. In particular, we will retain
only those configurations of angles and
masses for which, at a given value of $\sqrt{s}$, the total numbers of
$\epem \ra Z h_1$ and (separately) $Z h_2$ events are both 
less than 50 for an integrated luminosity
of $500 \fbi$. In Fig.~\ref{h1h2_cpodd_contour} we
show contour plots for the minimum value of the pair
production cross section, $\sigma(e^+e^- \rightarrow h_1 h_2)$,  
as a function of
Higgs boson masses at $\sqrt{s}=350$, 500, 800 and 1600 GeV. With 
integrated luminosity of $500 \fbi$, a large number of events (large
enough to allow for selection cuts and experimental efficiencies) 
is predicted for all the above energies 
over a broad range of Higgs boson masses.
If 50 events before cuts and efficiencies prove adequate, one can
probe reasonably close to the kinematic boundary defined
by requiring that $m_{h_1}+\mz$, $m_{h_2}+\mz$ and $m_{h_1}+m_{h_2}$
all be less than $\sqrt s$.

\item{(ii)} $m_{h_1}+\mz<\sqrt s$, $m_{h_1}+m_{h_2},m_{h_2}+\mz>\sqrt{s}$:

In this case, if $C_1$
is small the sum rules (\ref{yuksr}) imply that Yukawa
couplings may still allow detection of the $h_1$. We illustrate this in
Fig.~\ref{fig_limits} by plotting the minimum and maximum values of
$\sigma(e^+e^- \rightarrow f\bar{f}h_1)$ for $f=t,b$
as a function of the Higgs boson mass, where the scan
over the mixing angles $\alpha_1$, $\alpha_2$ and
$\alpha_3$ at a given $\tanb$ is constrained by present experimental
constraints and by the requirement that 
fewer than 50 $Zh_{1}$ events are predicted for 
$\sqrt{s}=500$ GeV and $L=500\fbi$.
[The results are essentially independent of $m_{h_2}$ (and $m_{h_3}$)
for $m_{h_1}+m_{h_2}>\sqrt s$.] For comparison, the full
lines are the cross sections for $e^+e^-\rightarrow f\bar{f}\ha$ in the
CPC model with $\mha=m_{h_1}$. From Fig.~\ref{fig_limits},
we conclude that if $m_{h_1}$ is not large there will be sufficient events in 
either the $b\bar{b}h_1$ or the
$t\bar{t}h_1$ channel (and perhaps both) to allow its discovery.
Clearly the most pessimistic scenario is one with $1\lsim\tanb\lsim 10$
and minimal $t\anti t h_1$ cross sections; taking 50 events (before
cuts and efficiencies) as the criteria, and assuming $L=500\fbi$
at $\rts=500\gev$,
one could only detect an $h_1$ with $C_1\sim 0$ if $m_{h_1}\lsim 70\gev$
(just as for the $\ha$ in the CPC model). A $\rts=1\tev$ machine
would considerably extend this mass reach.

Several points are worth noting:
\bit
\item
For a given $\tanb$ value, the $C_1\sim 0$ cross sections of
Fig.~\ref{fig_limits} exhibit two important features. (a)
The minimal and maximal $b\anti bh_1$ cross sections are almost equal,
for a given $\tanb$ value,
and are essentially the same as the $b\anti b\ha$ cross section
in the CP-conserving two-doublet model. (b) The minimal $t\anti th_1$
cross section and the $t\anti t\ha$ cross section are essentially equal.
\item
That the $C_1\sim 0$ 
cross sections should be related to the $\ha$ cross sections is not
altogether surprising given that
in the limit of $C_1\to 0$ the $h_1$ behaves like the $\ha$
in that it decouples from $ZZ$. However, to understand
why (for $C_1\sim 0$) the minimal and maximal $h_1$ cross sections and 
the $\ha$ cross section
are all numerically essentially the same in the $b\anti b$ 
final state, despite the fact that the $h_1$
possesses  non-zero $S$ and $P$ Yukawa couplings (and therefore is
not a genuine pseudoscalar) requires more discussion.
First, we note that, for $C_1\to 0$, Eq.~(\ref{yuksr}) implies
\beq
\yts \to (\hat P_{\ha}^t)^2\,,\quad \ybs \to (\hat P_{\ha}^b)^2\,,
\label{speclim}
\eeq
where $P_{\ha}^{t,b}$ are the $t$ and $b$ couplings of the $\ha$
in the CP-conserving version of the 2HDM.
Second, we note that in Eq.~(\ref{diffcross}) only
$G_{1,2}=(S_i^f)^2g_{1,2}$ and $F_{1,2}=(P_i^f)^2f_{1,2}$
[where $g_{1,2}$ and $f_{1,2}$ are functions of kinematic variables only,
defined by Eqs.~(\ref{gfuns}) and (\ref{ffuns}) in the Appendix]
remain non-zero as $C_1\to 0$ (see Appendix), implying in rough notation:
\beq
{d\sigma(\epem\to f\anti f h_1)\over d\phi}\sim (S_1^f)^2(Ag_1+Bg_2)+(P_1^f)^2(Af_1+Bf_2)
,,
\label{rougheq}
\eeq
where $A$, $B$, $f_{1,2}$ and $g_{1,2}$ are all positive
and $\phi$ denotes a point in phase space.
Thirdly, it is easily verified that $g_1-f_1$ and $g_2-f_2$ are both of order
$m_f^2/s$ and thus differ very little in the case of the $b\anti b$
final state. As a result, inserting the $C_1\to 0$ limit of
Eq.~(\ref{speclim}) into Eq.~(\ref{diffcross}) [or Eq.~(\ref{rougheq})] 
implies that the minimal and maximal
values of $\sigma(\epem\to b\anti bh_1)$ are essentially the same
and that both are very nearly equal to $\sigma(\epem\to b\anti b \ha)$.

\item
Next, we would like to understand why the minimum $t\anti t h_1$
cross section is obtained by taking $S_1^t\sim 0$, equivalent to
[see Eq.~(\ref{speclim})] $(P_1^t)^2\sim (P_{\ha}^t)^2$. Referring
to Eq.~(\ref{rougheq}), we see that this will be the case if
$\int (Ag_1+Bg_2) d\phi>\int (Af_1+Bf_2) d\phi$, as is easily
verified.
\item
We note that the minimum cross section values 
would be altered if the scan over the $\alpha_i$
is not restricted by requiring small $C_1$. 
In particular, if one observes $Zh_1$ events and finds $|C_1|\sim 1$,
then, as outlined earlier, both $\yts$ and $\ybs$ will
be of order unity, approaching 1 exactly if $\tanb$ is either very large
or very small. This implies minimum cross sections values similar
to the $\tanb=1$ $f\anti f\ha$ cross sections. Thus, 
at a $\rts=500\gev$ machine with integrated
luminosity of order $L=500\fbi$, it would
almost certainly not be possible to use $b\anti b h_1$
production to measure the $h_1$'s $b\anti b$ coupling 
and, if $m_{h_1}$ is
significantly above $70\gev$, it would also be difficult to measure its
$t\anti t$ Yukawa coupling. Of course, increasing the $\rts$
will extend the range of $m_{h_1}$ for which the $t\anti t h_1$ process
will have a useful rate.
\eit

\end{description}

We note that, even if $L=500\fbi$ cannot be achieved in a single
year of operation at $\rts\sim 500\gev$, 
one can envision accumulating such an integrated
luminosity over a period of several years.
For $\rts\sim 1\tev$ and above, our results may be conservative
given that the $\epem$
collider will very probably be designed to have a yearly integrated
luminosity that scales with energy like $s$.

\section{The two-doublet + one-singlet (2D1S) Higgs sector model}

We do not go into the details of the most general Higgs potential
for the 2D1S model, but simply state the well-known fact that
explicit or spontaneous CP violation is entirely possible for
a 2D1S Higgs sector.
The primary change relative to the formalism given for the two-doublet
model is that the $R$ matrix is extended to a $5\times 5$ matrix.
The formulae for the couplings of a given physical eigenstate $h_i$
to $ZZ$ and to the quarks remain unchanged relative to the two-doublet
case, being entirely determined by $R_{i1}$, $R_{i2}$ and $R_{i3}$
in the basis where 
\beq
(\varphi_1,\varphi_2,\varphi_3,\varphi_4,\varphi_5)\equiv
\sqrt 2(\mbox{Re}\phi_1^0, \, \mbox{Re}\phi_2^0, \,
 s_\beta\mbox{Im}\phi_1^0-c_\beta\mbox{Im}\phi_2^0,\mbox{Re}N,\mbox{Im}N) \,,
\label{sphi}
\eeq
with $N$ being the singlet Higgs field. In general, the only constraints
on the parameters of the model are that $R$ must, as before, be
an orthogonal matrix and the masses-squared 
of the physical Higgs eigenstates must be non-negative. Physically,
this means that we can have two light Higgs bosons that reside entirely
within the singlet sector and therefore do not couple to either
quarks or gauge bosons. As a result, one can only
guarantee discovery of a neutral Higgs boson if at least three of the five
physical states are light.
Further, we shall show that this guarantee is possible only
by employing the Yukawa radiation processes.
No statement will be possible for just one or two light Higgs bosons.

We begin by focusing on the generalization of the Yukawa sum rules
to the 2D1S case. Starting from Eq.~(\ref{yuk}) (which still applies),
one finds 
\begin{eqnarray}
  \label{finalsrp}
  \sin^2\beta [(\hat{S}^t_i)^2 + (\hat{P}^t_i)^2]
+ \cos^2\beta [(\hat{S}^b_i)^2 + (\hat{P}^b_i)^2]=
R_{i1}^2+R_{i2}^2+R_{i3}^2\equiv R_i^2\,,                                               
\end{eqnarray}
where $R_i^2$ is a measure of the extent to which $h_i$ resides
in the two-doublet portion of the Higgs sector.
We will refer to $R_i^2$ as the two-doublet content of $h_i$.
In the 2HDM model $R_i^2=1$ ($i=1,2,3$) was automatic by virtue of 
the orthogonality of $R$ and its $3\times 3$ dimensionality. 
However, in the present case
$R_i^2=R_{i1}^2+R_{i2}^2+R_{i3}^2=1-R_{i4}^2-R_{i5}^2$ could be zero if the
$h_i$ Higgs boson 
resides entirely in the singlet sector ($R_{i4}^2+R_{i5}^2=1$).
We only know that after summing over all the physical Higgs
bosons we must get the full two-doublet content: $\sum_{i=1}^5 R_i^2=3$.
Results analogous to the $C_i=0$ limits of Eqs.~(\ref{yuksr}) can also
be obtained. For $C_i= 0$,
\beq
(\hat S_i^t)^2+(\hat P_i^t)^2=\left({\cos\beta\over\sin\beta}\right)^2R_i^2\,,
\quad
(\hat S_i^b)^2+(\hat P_i^b)^2=\left({\sin\beta\over\cos\beta}\right)^2R_i^2\,.
\label{tbs}
\eeq
Note that both could be zero for a pure singlet $h_i$. Summing over two
Higgs bosons does not help, since both Higgs
could reside entirely in the singlet
sector. However, if we sum over three Higgs bosons (we use $i=1,2,3$
in what follows), one finds
\beq
\sum_{i=1,2,3}R_i^2=1+(R_{44}^2+R_{45}^2+R_{54}^2+R_{55}^2)\geq 1\,.
\label{sumiii}
\eeq
In the worst case, $R_{44}^2=R_{45}^2=R_{54}^2=R_{55}^2=0$, \ie\ the singlet
Higgs field $N$ is entirely contained in the three light Higgs bosons.
The two most important implications of these results are the following.
\begin{enumerate}
\item 
Eq.~(\ref{finalsrp}) implies that
our ability to observe a Yukawa radiation process
and measure either the $b\anti b$ or the $t\anti t$ Yukawa
coupling of a Higgs boson $h_i$ is determined by its two-doublet
content, $R_i^2$. For substantial two-doublet content, 
and $m_{h_i}+2m_t<\sqrt s-\Delta$, 
we are guaranteed that at least one of these two Yukawa
couplings will be measurable.
\item
If there are three light Higgs bosons (light being defined by
$m_{h_i}+2m_t<\sqrt s-\Delta$), and two have small Yukawa
couplings, then Eq.~(\ref{sumiii}) implies that
at least one of the Yukawa couplings of the third
will be large enough to detect the Higgs boson in association
with $b\anti b$ or $t\anti t$.

Of course, the Yukawa couplings (squared) could be apportioned more
or less equally among the three light Higgs bosons, in which case
observation of a Yukawa radiation process of any one of the three would
require substantially more luminosity than if the two-doublet content resides
primarily in just one of the three.
\end{enumerate}

The generalization to more singlets is clear.  Each singlet field
introduces two more physical neutral Higgs bosons. At least
$1+2N_{\rm singlet}$ of the neutral Higgs bosons must be light in order
to guarantee that $\sum_{i=1}^{1+2N_{\rm singlet}} R_i^2\geq 1$,
implying definite opportunity for observing at least one in $t\anti t h_i$
or $b\anti b h_i$ associated production.

Let us now consider the $Zh_i$ and $h_ih_j$ processes. We wish
to determine how many of the 2D1S neutral Higgs bosons must be light in order
that we are guaranteed to find at least one in either Higgs-strahlung
or Higgs pair production. The crucial ingredient for
obtaining the necessary sum rule is 
the unitarity sum rule for $ZZ\to h_i h_j $ as given in 
the Appendix of Ref.~\cite{ghw}. In applying this sum rule it
is crucial to note that
the $ZZ$-Higgs-Higgs coupling only receives contributions from
the fields in the doublet sector. Thus, in the basis defined
by Eq.~(\ref{sphi}), these interactions
have the form $ZZ(\varphi_1^2+\varphi_2^2+\varphi_3^2)$ 
times the standard $g^2/(2c_W^2)$ factor.
There are no $ZZ\varphi_4^2$ or $ZZ\varphi_5^2$ interactions.
After diagonalizing, the
$ZZh_ih_j$ coupling coefficient is given [see Eq.~(\ref{zzhhcoup})] by 
$X_{ij}\equiv R_{i1}R_{j1}+R_{i2}R_{j2}+R_{i3}R_{j3}$.
In particular, $X_{ii}=R_i^2$, the two-doublet content of $h_i$
defined earlier. Using our present notation, Eq.~(A18) of Ref.~\cite{ghw}
becomes
\beq
C_iC_j+\sum_{k\neq i} C_{ik}C_{jk}=X_{ij}\,,
\label{correctsrgeneral}
\eeq
which for $i=j$ yields
\beq
C_i^2+\sum_{k\neq i} C_{ik}^2=R_i^2\,.
\label{correctsr}
\eeq
Let us define
\beq
W_{1234}\equiv 
C_1^2+C_2^2+C_3^2+C_4^2+C_{12}^2+C_{13}^2+C_{14}^2+C_{23}^2+
C_{24}^2+C_{34}^2\,.
\label{wdef}
\eeq
Using Eq.~(\ref{correctsr}) and summing over $i=1,2,3,4$
and over $i=1,2,3,4,5$, one obtains
\bea
&&W_{1234}+\sum_{i,j=1,\ldots,5:~i>j}C_{ij}^2=\sum_{i=1}^4R_i^2=3-R_5^2\,,
\label{sr1}\\
&&\sum_{i=1}^5C_i^2+2\sum_{i,j=1,\ldots,5:~i>j}C_{ij}^2
=\sum_{i=1}^5R_i^2=3\,,
\label{sr2}
\eea
respectively, where we also used $C_{ik}^2=C_{ki}^2$.
Unitarity for $ZZ\to ZZ$ scattering
and for other vector boson $VV\to VV$ processes 
requires that $\sum_{i=1,5}C_i^2=1$.
Inserting this into Eq.~(\ref{sr2}) implies that 
$\sum_{i,j=1,\ldots,5:~i>j}C_{ij}^2=1$. Inserting this latter result into
Eq.~(\ref{sr1}) yields $W_{1234}=2-R_5^2$ which must be $\geq 1$
by virtue of the fact that $R_5^2\leq 1$ is required by orthogonality of $R$.
In words, $W_{1234}\geq 1$ implies that if there are four
Higgs bosons that are sufficiently light 
that all the $Zh_i$ and $h_ih_j$ production processes are kinematically
allowed (and not significantly phase-space suppressed),
then at least one of these Higgs bosons must be seen in Higgs-strahlung
or a pair of Higgs bosons must be seen in pair production. Three light
Higgs bosons are not enough.  In particular, analogous procedures
to those sketched above yield the result
\beq
W_{123}\equiv
C_1^2+C_2^2+C_3^2+C_{12}^2+C_{13}^2+C_{23}^2=\sum_{i=1}^3R_i^2-1+C_{45}^2\,.
\eeq
Since we are only guaranteed that $\sum_{i=1}^3R_i^2\geq 1$ and
since $C_{45}$ could be quite small even when 
$\sum_{i=1}^3R_i^2= 1$, there is no lower bound to $W_{123}$ and we
cannot be certain of finding at least one Higgs boson 
in Higgs-strahlung or Higgs pair production in the case
that only three are light.~\footnote{Note: these results correct
the erroneous result for this case given in Ref.~\cite{gghk}.}
Thus, if only three neutral Higgs bosons of the 2D1S model are light,
searching for the Yukawa radiation processes is required in order
to guarantee that we will find at least one.

Once again, the generalization of the above considerations
to a CP-violating Higgs sector
with one-doublet and more than one singlet is obvious.
At least $2+2N_{\rm singlet}$ of the neutral Higgs bosons must
be light in order to be certain
that at least one of them will be produced at a significant
rate in either Higgs-strahlung or Higgs pair production.

\section{Discussion and conclusions}

We have derived a crucial new sum rule, Eq.~(\ref{yuksr}),
relating the Yukawa and Higgs-$ZZ$ couplings of a general CP-violating
two-Higgs-doublet model. This sum rule has two important
implications. First, it says that if the $ZZh$
coupling of a neutral Higgs boson is small, then its $t\anti t h$
or $b\anti b h$ Yukawa coupling must be substantial. This means
that any one of the three neutral
Higgs bosons that is light enough to be produced in $\epem\to t\anti t h$
(implying that $\epem\to Zh$ and 
$\epem\to b\anti b h$ are also kinematically allowed) will
be found at an $e^+e^-$ linear collider of sufficient luminosity.
In particular, if mixing angles and Higgs masses 
are such that a light Higgs boson cannot be observed via the $Zh$ 
Higgs-strahlung process, then it is guaranteed to be found 
via Yukawa-coupling-induced radiation from top or bottom quarks.
Second, for an $h$ that is observed in the $Zh$ final state
but also light enough to be seen in $t\anti t h$ and, by implication,
$b\anti b h$, this same sum rule can be used to show that measurement of
at least one of its third-family Yukawa couplings will be possible
(the required luminosity depending on the amount
of phase space suppression in the $t\anti t h$ channel). 
Of course, in the experimental
analysis one must be careful to not exclude the Yukawa radiation processes
by placing restrictive invariant mass constraints on the $f\bar{f}$ system,
\eg, $M_{f\anti f}\sim m_Z$. 

We have also extended to high energies the quantitative analysis of
a previously derived sum rule, Eq.~(\ref{oldsr}). This 
latter sum rule implies that
if any two of the three neutral Higgs bosons of the CP-violating
2HDM are light enough that $Zh_1$, $Zh_2$ and $h_1h_2$
production are all kinematically allowed (and not phase space suppressed),
then at least one of these processes will be observable,
regardless of the mixing structure of the neutral Higgs sector.
For planned luminosities, the predicted cross sections are
such that discovery of
one or both of the Higgs bosons will be possible even
rather close to the relevant kinematic boundary in the $m_{h_1}$ - $m_{h_2}$
mass plane.

We have also considered the general CP violating 
two-doublet + one-singlet Higgs sector model.  In this case, 
we find that if only one or two of the neutral Higgs
bosons are light then both could be primarily singlet and, therefore,
undetectable in Higgs-strahlung, Higgs pair 
production and Yukawa radiation processes. However,
there are two important guarantees. (a) If there are three light 
neutral Higgs bosons, then we are guaranteed to detect at least one
in Yukawa radiation processes. 
(b) If there are four light neutral Higgs bosons we are guaranteed to detect
one or two in Higgs-strahlung or Higgs pair production; but,
there is no such guarantee for just three light Higgs bosons.
Guarantee (a) requires that all $t\anti t h_i$ ($i=1,2,3$)
(and by implication all $b\anti b h_i$) processes have
substantial phase space.
Guarantee (b) requires that all four $h_i$ be light enough that the
$Zh_i$ and $h_ih_j$ ($i,j=1,2,3,4$, $i\neq j$) processes all have
substantial phase space.   Thus, for extensions of the two-doublet
Higgs sector that include one or more singlet Higgs fields,
it could happen that observation of a Higgs boson 
at an $\epem$ collider of limited energy will
only be possible by looking for Higgs production
in association with bottom and top quarks.

\vspace{1.5cm}
\centerline{\bf Acknowledgments}
\vspace{.5cm} This work was supported in part by the Committee for
Scientific Research (Poland) under grants No. 2~P03B~014~14, No.
2~P03B~030~14, by Maria Sklodowska-Curie Joint Fund II
(Poland-USA) under grant No. MEN/NSF-96-252, by the U.S.
Department of Energy under grant No. DE-FG03-91ER40674
and by the U.C. Davis Institute for High Energy Physics.

\vspace{1cm}
\noindent{\Large\bf Appendix }

\bigskip

Consider production of the Higgs 
boson $h_i$ in association with a fermion pair $f\bar f$ in $\epem$
collisions, \ie\ $\epem\to f\bar{f}h_i$. 
Note that the diagram with Higgs pair production
requires summation over virtual Higgs bosons $h_j$ and $h_k$, where
$i,j,k$ are permutations of $1,2,3$.
The differential cross section is given by Eq.~(\ref{diffcross})
with $F_i$ and $G_i$ as given below.

For a short hand notation, we introduce 
$h_j=m_{h_j}^2/s$, $\gamma_j=\Gamma_{h_j}^2/s$, (j=1,2,3)  
and $f=m_f^2/s$. The reduced 
energy of the observed Higgs boson $h_i$ is denoted by 
$x=2 E_{h_i}/\sqrt{s}=2-x_1-x_2$; we also define
$x_{12}=(1-x_1)(1-x_2)$.
In the formulae below, $Z$ and $h_j$ widths are included
in terms corresponding to $Z$ and $h_j$ decay to the $f\bar f$ pair. 

The functions $G_1$ and $G_2$  describe the 
$h_i$ Higgs boson radiation off the fermions
due to the scalar couplings,
\bea
      G_1&=&\frac{(S_i^f)^2}{4\pi x_{12}}\left[x^2-h_i(\frac{x^2}{x_{12}}+2 
    (x -1-h_i))
     +2 f \left(4(x-h_i)+\frac{x^2}{x_{12}} (4 f-h_i+2)\right)\right],\non \\
      G_2& =& -\frac{2 (S_i^f)^2}{4\pi x_{12}}
         \Biggl[  x_{12} (1+x)-h_i(x_{12}+8f+2x-2h_i) 
\non\\&\phantom{=}&\phantom{-\frac{2 (S_i^f)^2}{4\pi x_{12}}}\,
     +3f x \left(\frac{x}{3}+4+\frac{x}{x_{12}}(4f-h_i)\right)\Biggr].
\label{gfuns}
\eea
whereas the $F_1$ and $F_2$ terms arise from the pseudoscalar couplings,
\bea
      F_1&=& \frac{(P_i^f)^2}{4\pi x_{12}} \left[x^2-h_i
       (\frac{x^2}{x_{12}}(1+2f)+2 x-2-2h_i)\right], \non \\
      F_2&=& \frac{2(P_i^f)^2}{4\pi x_{12}} \left[ (2h_i-x_{12}) 
         (1+x-h_i)-2h_i (1+2f)
      - \frac{f x^2}{x_{12}} (x_{12}-3h_i)\right].
\label{ffuns}
\eea

The terms $G_3$ and $G_4$ account for the emission of the Higgs boson
(only its $CP=1$ component) from the $Z$-boson line:
\bea
      G_3&=&\frac{2 g_{ZZh_i}^2}{4\pi  (p^2+z\gamma_z)}
     \left[f(4h_i-x^2-12 z)+\frac{f}{z}(4h_i-x^2) (x-1-h_i+z)\right], \non \\
      G_4&=&\frac{2z g_{ZZh_i}^2}{4 \pi  (p^2+z \gamma_z)}
        \left[h_i+x_{12}+2-2 x+4f\right], 
\eea
where the reduced propagator of the off-shell $Z$-boson has been 
denoted by $p=x-1-h_i+z$.

The interference between the radiation amplitudes off the 
fermion and the $Z$-boson lines is included in the
$G_5$ and $G_6$ terms:~\footnote{Due to a different convention regarding 
the sign of the $g_{ZZH}$ coupling, our $G_5$ and $G_6$ have opposite signs 
to those in Eq.~(9) of~\cite{dkz}.}
\bea
      G_5&=&\frac{S_i^f g_{ZZh_i}}{4\pi}\, \frac{4x  m_f}{x_{12} m_Z}
        \,\frac{p}{p^2+z \gamma_z}
        \Bigl[(x_{12}-h_i)(x-1-h_i) \non \\
       &+&  f(12 z-4h_i+x^2)-3 z h_i+6z x_{12}/x\Bigr],\non \\
      G_6&=&\frac{S_i^f g_{ZZh_i}}{4\pi}\, \frac{4z  m_f}{x_{12} m_Z}
        \,\frac{p}{p^2+z \gamma_z}
       \left[x (h_i-4f-2)  -2x_{12}+x^2\right].
\eea

Finally, the contributions from the Higgs pair $h_i h_j$ and 
$h_i h_k$ production diagrams 
(with subsequent $h_j$ and $h_k$ decays to fermion pairs) and from 
their interference  with the $h_i$ radiation off the 
fermion and the $Z$-boson lines  are collected in $G_7$, $G_8$,   
 $F_3$ and $F_4$ as follows:~\footnote{We correct 
some typos in  \cite{dkz}. The last term 
for $G_7$ in Eq.~(16) of~\cite{dkz} should have the opposite sign, 
\ie\ $-2a_fg_{ffH^i}$ should read $+2a_fg_{ffH^i}$. 
In Eq.~(18),  an overall factor of 4 multiplying $F_3$
is missing and $g_{ffA}$ should read 
$g_{ffH^i}$, and for $F_4$ a factor of 2 is missing. We thank 
S. Dawson and M. Spira for help in clarifying this point.}
\bea
      F_3&=& \frac{1}{2\pi} 
               (x-1-h_i+4f)(4h_i-x^2)
       \frac{(P_k^f C_{ik} u_j+P_j^f C_{ij} u_k)^2}
        {(u_j^2+h_j \gamma_j)(u_k^2+h_k \gamma_k)} ,\non \\
      F_4&=& - \frac{P_i^f}{\pi} 
     \left[\frac{S_j^f C_{ij} u_j}{u_j^2+h_j \gamma_j}
       +\frac{S_k^f C_{ik} u_k}{u_k^2+h_k \gamma_k}\right]\non \\
       &\times& \frac{x}{x_{12}}\left[(x_{12}-h_i)(1-x+h_i)
       +f(4h_i-x^2)\right], \\
      G_7&=&\frac{4h_i-x^2}{2\pi}\Biggl[   (x-1-h_i) 
                  \frac{(P_k^f C_{ik} u_j+P_j^f C_{ij} u_k)^2}{
       (u_j^2+h_j \gamma_j)(u_k^2+ h_k \gamma_k)}\non \\
         &+ &4 s_W c_W  \frac{2m_f}{m_Z} a_f g_{ZZh_i}
     (\frac{P_j^f C_{ij} u_j}{u_j^2+h_j \gamma_j}
     +\frac{P_k^f C_{ik} u_k}{u_k^2+h_k \gamma_k})\Biggr],\non \\
       G_8&=& \frac{S_i^f}{\pi}
         \left[\frac{P_j^f C_{ij} u_j}{u_j^2+h_j \gamma_j}
         +\frac{P_k^f C_{ik} u_k}{u_k^2+h_k \gamma_k}\right]\non \\
      &\times& \frac{x}{x_{12}}\left[(x_{12}-h_i) 
      (x-1-h_i)-f (4h_i-x^2)\right].
\eea
In the above expressions, terms of order  
$\gamma_i$, (i=1,2,3) in the numerator have been neglected. 
The scaled  propagator of the virtual Higgs boson $h_j$ 
  has been abbreviated by 
(for the virtual $h_k$ boson, replace $j\ra k$)

\beq 
      u_j=x-1-h_i+h_j
\eeq

If the Higgs  and $Z$ boson widths
are neglected then the above expressions reduce to those given in
Ref.~\cite{dkz} with the exception 
that our $G_7+ 4s_W c_W a_f G_8$
becomes $G_7 $ of \cite{dkz}.

\end{document}